# Enabling large language model agents for high-fidelity many-body perturbation theory calculations

Benran Zhang[1], Byeongchan Lee[2,3], Zien Zhu[1], Chih-En Hsu[1,4], Zhenfa Zheng[1], Hung-Chung Hsueh[4], Young Woo Choi[2,3], Mauro Del Ben[5], Zhenglu Li[1,a)]

[1] Mork Family Department of Chemical Engineering and Materials Science, University of Southern California, Los Angeles, CA, USA

[2] Department of Physics, Sogang University, Seoul, South Korea

[3] Center for Nano Materials, Sogang University, Seoul, South Korea

[4] Department of Physics, Tamkang University, Tamsui, New Taipei, Taiwan

[5] Applied Mathematics and Computational Research Division, Lawrence Berkeley National Laboratory, Berkeley, CA, USA

[a)] Author to whom correspondence should be addressed: zhenglul@usc.edu

**Abstract**

Many-body perturbation theory (MBPT) methods, including the GW approximation, the GW plus Bethe-Salpeter equation (GW-BSE) method, and GW perturbation theory (GWPT), among others, provide accurate first-principles descriptions of excited-state properties and multiparticle interactions of materials. However, their use remains limited due to complex multi-software workflows, interdependent physical parameters, and substantial theoretical expertise. Large language model (LLM) agents, which couple a language model to software tools, show potential in automating computational tasks, but MBPT remains an unaddressed challenge. We show that LLM agents tested in our work, including frontier models such as Opus 4.7 and GPT-5.5, do not reliably construct MBPT workflows in one attempt, suggesting that correct setup requires expert knowledge and consistency that current models lack. Across tested model-harness combinations, out-of-the-box agents average 54%, 36%, and 19% on GW, GW-BSE, and GWPT workflow benchmarks, with strong model dependence and large run-to-run variation. Errors typically originate from cross-stage inconsistencies that surface only several stages after they are introduced. To address these challenges, we develop AI-Donkey, an agentic framework that combines expert-curated knowledge with deterministic execution to enable high-fidelity GW, GW-BSE, and GWPT calculations and analyses. With AI-Donkey, all tested agents, from frontier proprietary models to open-weight models, produce complete, runnable MBPT workflows with no observed run-to-run variation. We further demonstrate autonomous calculations and post-processing capabilities on one-dimensional, two-dimensional, and bulk materials, including the many-body enhancement of Fröhlich electron-phonon coupling in polar semiconductors. Our approach provides a general framework for knowledge engineering to enable reliable agentic workflows in computational materials science.

## 1. Introduction

Excited-state properties such as quasiparticle band gaps and optical responses are critical properties that define the usefulness of a wide range of functional and energy materials[1–3]. Accurately predicting these excited-state properties can be achieved through many-body perturbation theory (MBPT) methods such as the GW approximation for accurate quasiparticle energies[2–5], GW-Bethe-Salpeter equation (GW-BSE) for optical properties[6,7], and GW perturbation theory (GWPT) for electron-phonon coupling[8,9]. Over the past decades, dedicated MBPT software packages and functionalities have been developed, including BerkeleyGW[10–12], yambo[13], WEST[14], ABINIT[15], exciting[16], and MOLGW[17], among others. Building on top of these packages, recent algorithmic advances and exascale optimizations have enabled simulation of excited-state materials properties at an unprecedented scale of system size and computational performance[11,12,18,19].

Yet, applying these methods in practice remains technically demanding. For instance, the standard GWPT[8] electron-phonon coupling calculation workflow[9] spans across 4 different software packages, namely, ABINIT[15], BerkeleyGW[10], Wannier90[20], and EPW[21], and consists of 20 individual steps[9]. Furthermore, parameter setups in the workflow are interdependent and deeply associated with the underlying theory and implementation details, making the decisions highly non-trivial. Setting up such a calculation requires not only familiarity with the input structure of each code, but also an understanding of how data, file formats, and the fundamental physics propagate across stages. This complexity in both the formalism and multi-code workflow has, to some extent, limited the wide adoption of MBPT methods despite their excellent predictive power.

Recent advances in large language models (LLMs) offer a potential route to lowering this barrier through LLM agents, where an LLM is coupled with a software harness that enables it to access information, invoke tools, and interact with computational environments to accomplish specified tasks. Early efforts focusing on chemistry and materials science fields combine LLM agents with domain-specific tools to bridge the gap between computational and experimental work[22], and to enable autonomous research[23]. More recently, these efforts have extended to the first-principles computation of materials properties, where LLM agent systems autonomously set up density functional theory (DFT) calculations[24–30]. These systems have been implemented as hierarchical multi-agent systems[24,25] and model context protocol (MCP) interfaces[26,27], demonstrating expert-level capabilities for automating ground-state DFT workflows.

However, these successes do not easily generalize to MBPT methods. DFT workflows typically reside within a single code with well-documented inputs that involve few cross-stage dependencies. MBPT workflows, on the other hand, span multiple codes with strict cross-stage consistency requirements. Inconsistent parameters often propagate through the workflow chain and cause downstream errors that are difficult to identify. Furthermore, MBPT is a rapidly advancing field, with new methods and capabilities emerging faster than they are captured in LLM training data. LLM agents benchmarked in this work, including frontier models such as Opus 4.7 and GPT-5.5, do not reliably construct MBPT workflows in one attempt, suggesting that correct setup requires expert knowledge and consistency that current models lack. Extending agentic automation to MBPT is nevertheless a critical step. It would broaden access to the predictive power of MBPT and establish a framework for agentic automation of complex, multi-code workflows.

To address this gap, we develop AI-Donkey, a portable, multi-code-compatible, and harness-agnostic agentic framework that enables autonomous end-to-end GW, GW-BSE, and GWPT calculations, centered around the MBPT software package BerkeleyGW[10–12]. AI-Donkey is built with an expert-curated knowledge layer, a set of portable agentic protocols, and a complete suite of command line interface (CLI) tools (Fig. 1), and is interfaced with Quantum ESPRESSO[31], BerkeleyGW[10], ABINIT[15], EPW[21], and Wannier90[20]. Its modular design allows for additional codes and MBPT workflows to be seamlessly integrated with its core architecture. AI-Donkey enables complex MBPT workflows, as demonstrated with various closed- and open-weight LLM agents, while reliably enforcing cross-stage consistency and end-to-end correctness. In this work, we perform calculations on several different supercomputing platforms, where we need to delegate tasks related to computational job submission to humans to follow cluster queue policy and restrictions; in a computing environment with full permission, full autonomy can be straightforwardly realized. On GW, GW-BSE, and GWPT evaluations, tested out-of-the-box LLM agents (i.e., Claude Code, Codex, open-weight harness) average 54%, 36%, and 19% on static benchmark scores and quality varies substantially between repeated trials, whereas AI-Donkey produces runnable, consistent workflows that score full marks across all repeated model-harness combinations benchmarked in this work. We further demonstrate the capabilities of the AI-Donkey framework by computing excited-state properties of one-dimensional (1D), two-dimensional (2D), and bulk systems autonomously. Additionally, we showcase the autonomous computation of the many-body enhancement of Fröhlich electron-phonon coupling of a polar semiconductor using GWPT. By enforcing consistent, deterministic execution and leveraging the dynamical reasoning ability of LLMs, AI-Donkey significantly lowers the execution barrier and enables automated capabilities to perform advanced MBPT calculations.

## 2. Results

AI-Donkey is designed as a stand-alone knowledge and execution layer that interfaces between existing LLM agents and first-principles software packages to enable automated MBPT workflows. Users interact with AI-Donkey through an agentic LLM harness of their choice, such as Claude Code or Codex, using natural-language interactions. The LLM agent then invokes AI-Donkey components to plan, execute, and validate MBPT workflows, as schematically shown in Fig. 1. For long-horizon tasks such as convergence tests that span multiple LLM agent sessions, AI-Donkey supports iterative research loops in which a high-level objective is decomposed into detailed deliverables, executed through deterministic tools, and evaluated against target criteria.

At its core, AI-Donkey has four main components consisting of agentic workflow protocols, retrieval-augmented generation (RAG), expert-curated skills, and command-line interface (CLI) tools (Fig. 1). The *workflow protocols* are structured natural-language procedures that guide the agent through a calculation in ordered steps. Each step is mapped to the relevant skills and CLI tools, making the protocols modular and extensible. The *RAG component* provides semantic retrieval of expert knowledge, allowing the LLM agent to ground decisions in curated domain expertise. The *expert-curated skills* are modular units of procedural knowledge, each corresponding to a module in the underlying code. Each skill documents the module's purpose, required inputs and outputs, key physical choices, common failures, and corresponding CLI invocations. Because skills are modular, LLM agents can assemble them on the fly into coupled workflows, such as convergence studies. The *CLI layer* enforces deterministic input generation, dependency linking, and validation through a unified Python interface. Rather than relying on the LLM to generate code-specific syntax and track file dependencies, the CLI serves as the source of consistency and robustness in workflow setup. With this four-component design, AI-Donkey decouples high-level reasoning and workflow navigation, which are delegated to LLMs, from syntax control, dependency management, and input generation, which are enforced deterministically through the CLI.

To evaluate AI-Donkey's ability to carry out robust autonomous MBPT workflows, we use silicon GW and GW-BSE calculations as canonical benchmarks. These workflows have well-defined inputs and outputs, while still requiring nontrivial workflow logic, cross-step dependency propagation, and physics-aware parameter choices. They therefore provide a controlled baseline for assessing whether an LLM agent can perform standard MBPT calculations autonomously.

We apply this benchmark across a diverse set of LLM agents to test their performance with and without AI-Donkey and to validate the robustness and portability of AI-Donkey. The evaluated systems include Claude Code with Haiku 4.5[32], Sonnet 4.6[33], and Opus 4.7[34]; Codex with GPT-5.4-mini[35] and GPT-5.5[36]; and a custom lightweight, generic agentic harness paired with the open-weight gpt-oss-120b[37] and Gemma-4-31B[38] models. The lightweight harness provides only basic tool-calling capabilities, including file reading and writing, shell execution, and directory listing and inspection (see supplementary material for details). Together, these LLM agents allow us to evaluate AI-Donkey across a broad range of model capabilities and harness designs.

Each LLM agent is evaluated under two knowledge conditions: with access to AI-Donkey and out-of-the-box without AI-Donkey. The AI-Donkey condition tests whether structured expert knowledge and deterministic execution tools enable reliable MBPT workflow construction, whereas the out-of-the-box condition tests whether a general-purpose agentic harness can infer and execute the same workflows using pretrained knowledge and native tool use alone. In the one-shot benchmark, each run is initialized with an identical prompt and proceeds without human intervention. Generated workflows are scored according to static workflow correctness, runnability of the generated calculation, and successful post-processing of the target deliverables. The details of the scoring rubrics are provided in the Methods section. To evaluate robustness, we repeat the benchmark 5 times and report the run-to-run sample standard deviation of the static workflow score.

The resulting one-shot scores and standard deviation analysis are shown in Fig. 2. Without AI-Donkey, LLM agent performance is strongly model-dependent and exhibits substantial run-to-run variability. Most LLM agents, including those based on open-weight models, generate largely correct DFT inputs, reflecting the broad availability of practical DFT knowledge in model pretraining data. Performance degrades substantially in the GW and GW-BSE stages and in cross-stage consistency, which we attribute to the limited MBPT knowledge available in pretraining data. Fig. 2B and 2D show that out-of-the-box harnesses have large standard deviation across repeated runs, making successful MBPT setup sensitive to model stochasticity. Here, the average static-benchmark scores in the out-of-the-box condition across the 7 tested LLM agents and 5 repeated trials are 53.6% and 36.0% for GW and GW-BSE tasks, respectively.

A representative failure of out-of-the-box LLM agents is the handling of symmetry in **k**- and **q**-point grids across the workflow. BerkeleyGW require either specifying the full **k**- and **q**- point grid across all steps or using a set of compatible symmetry-reduced mesh determined by the space group of the system. Opus 4.7 and Sonnet 4.6 violates requirements by hallucinating a symmetry-reduced grid that omits inequivalent **k**-points while duplicating others. These types of latent errors propagate the workflow and only surface during later computational steps when they are difficult to trace. Less capable models such as Haiku 4.5, GPT-5.4-mini, and open-weight models fail to provide a **k**- and **q-**point list for GW steps altogether. Symmetry settings are also not propagated coherently across stages. In all out-of-the-box GW-BSE workflows that specify a coarse-grid symmetry flag in the BSE stage, there is a mismatch between the actual wavefunction grid and the declared symmetry.

Providing AI-Donkey to the same LLM agents consistently remove the failures observed in the out-of-the-box conditions. Across all tested closed- and open-weight models, LLM agents identify the correct symmetry operations using dedicated symmetry-reduction tool through the CLI and reach the maximum score for both GW and GW-BSE tasks. Our analysis further shows that AI-Donkey substantially improves reproducibility across repeated generations with zero sample standard deviation in our tests, showing that workflow correctness is rigorously controlled by the structured knowledge and deterministic execution layer. In particular, the same level of success achieved by the open-weight LLM agents demonstrates that only minimal tool-calling capabilities are needed once expert MBPT workflow knowledge is supplied by AI-Donkey.

Having established AI-Donkey on standard GW and GW-BSE workflows, we next move to one of the most complex MBPT workflows, the GW perturbation theory (GWPT) calculations. The GWPT workflow requires coordination across 4 electronic-structure codes and involves extensive and strict file linking, setup consistency, and parameter-setting requirements between stages (Fig. 3A), including wavefunction gauge, symmetry, and commensurate momentum grids.

In Fig. 3B and 3C, we use the GWPT workflow for computing the phonon-induced electron linewidth of SiC[39] to benchmark Opus 4.7 with the Claude Code agentic harness and Gemma-4-31B with the lightweight harness. These two settings represent a frontier proprietary agentic environment and an accessible open-weight alternative, respectively. As a baseline, we evaluate each LLM agent out of the box, with no GWPT information provided. As of this work, GWPT is not a publicly released feature of BerkeleyGW, hence even the strong closed-weight proprietary models have not been trained with specific examples (at best with the published literature) and serves as an excellent blind test for such frontier models. We then provide access to the source code of all involved software packages, testing whether implementation-level information alone is sufficient for the models and agent to infer and reconstruct the workflow. Finally, we provide the full AI-Donkey framework, which encodes the GWPT workflow protocol, code-specific skills, parameter dependencies, and deterministic execution logic. Fig. 3B reports one of the scores of the above tests that was submitted to HPC resources for runnability testing, and Fig. 3C reports the sample standard deviation across 5 repeated trials.

The ablation study reveals that out-of-the-box LLM agents are insufficient for constructing runnable GWPT workflows. Without additional knowledge, LLM agents can identify parts of the DFT and DFPT setup correctly, likely reflecting pretraining knowledge of DFT workflows. However, both model-harness combinations fail on tasks that require specialized GWPT knowledge, missing cross-stage physical consistency and key GWPT-specific components. Providing source-code access leads to moderate improvement in the Opus 4.7 LLM agent and no distinguishable improvement in the Gemma-4-31B LLM agent. In the GWPT task, the model-average static-benchmark scores across the two tested LLM agents and 5 repeated trials are 18.6% and 29.6% in the out-of-the-box and with source-code access conditions, respectively.

As in the GW and GW-BSE workflows, symmetry is a major source of failure, and GWPT imposes additional requirements on the use of symmetry that directly affect the correctness of the results. The phonon **q**-grid in GWPT can only use spatial symmetry and not time-reversal symmetry, and the phonon **k**-grid is required to be the same as the one used in Wannierization. In addition, a single set of wavefunctions determines the gauge of the entire workflow, and its **k**-point mesh must be commensurate with and at least as dense as the one used in Wannierization. None of the 20 workflows generated without AI-Donkey, including those with source-code access, correctly infer these strict requirements of the **k**- and **q**- point grid. Opus 4.7 hallucinates the symmetry-reduced phonon **q**-grid and uses a phonon **k**-grid that is inconsistent with the Wannierization step, while Gemma-4-31B does not produce a phonon calculation at all. Even with source-code access, neither LLM agent can produce a usable GWPT workflow in a single attempt. The involved codes are large and complex, and implementation-level access alone does not expose the high-level workflow order, physical dependencies, or consistency constraints needed to prevent cross-stage inconsistencies. This highlights the necessity of human expert-curated skills and knowledge for newly developed methods.

Providing the full AI-Donkey framework enables both LLM agents to complete the GWPT workflow consistently. With AI-Donkey, all 10 runs reduce the phonon **q**-grid using only spatial symmetry and use a phonon **k**-grid consistent with the Wannierization. The deterministic CLI layer propagates dependencies, checks cross-stage consistency, and reduces the burden on the LLM to track long-range workflow constraints. At the same time, the expert-curated skills and workflow documentation present each workflow module in a compact, agent-readable form, allowing the agent to identify the correct workflow order, physical dependencies, and module-specific requirements. Both tested LLM agents achieve consistent GWPT workflow construction across the five repeated evaluations performed (Fig. 3C). Notably, AI-Donkey closes the gap between the frontier proprietary model–harness combination and the lightweight open-weight setup. These results show that AI-Donkey's advantage comes from structured knowledge engineering and deterministic execution.

We next demonstrate AI-Donkey's ability to produce and analyze MBPT results across a wide set of material systems. We select four representative materials spanning 1D, 2D, and bulk regimes. These systems are chosen to reflect the wide range of structures commonly encountered in excited-state materials research, while also imposing distinct physical and numerical requirements on the workflow. To succeed, the LLM agent must identify the appropriate treatment for each system's dimensionality and consistently propagate parameter choices through the workflow. These demonstrations show that AI-Donkey can generalize beyond canonical benchmarks to practical excited-state calculations.

In Fig. 4, we show results set up autonomously by Opus 4.7 with the Claude Code agentic harness using AI-Donkey. The LLM agent was instructed to set up complete, production-quality calculations for each material. The generated workflows, including input files, folder structures and file dependencies, were submitted to high-performance computing (HPC) platforms (the job submission actions are performed by humans). After the calculations completed, a post-processing agent was instructed to analyze the results and set up post-processing steps, including plotting and visualization.

We first evaluate AI-Donkey with a standard quasiparticle band-structure calculation for monolayer $MoSe_2$ with spin-orbit coupling (SOC) included (Fig. 4A). We then instruct the LLM agent to carry out GW-BSE calculation on the same material. As shown in Fig. 4B, the result obtained by the LLM agent shows the first exciton peak at 1.56 eV and is consistent with earlier literature references[40]. Furthermore, the post-processing agent separates the optical absorption spectrum of monolayer $MoSe_2$ into contributions from SOC-split A- and B-series excitons based on the band projection of the composing Bloch states in excitonic states. We next demonstrate the GW-BSE optical absorption spectrum calculation for graphene (Fig. 4C). Graphene is a computationally demanding system because the linear-dispersing Dirac cone requires a fine **k**- point sampling[41]. In both cases, the LLM agent correctly applied the system-specific workflow settings, generating runnable and physically correct workflows. AI-Donkey also supports specialized post-processing of excited-state properties. In Fig. 4D-F we show the real-space exciton wavefunctions of the (8,0) carbon nanotube, monolayer $MoSe_2$, and bulk diamond by fixing the hole position and plotting the corresponding electron distribution, respectively.

In addition, to test AI-Donkey's ability to support frontier MBPT research tasks, we instructed the LLM agent to investigate the many-body enhancement of Fröhlich electron-phonon coupling in the polar semiconductor SiC, following Ref. 39. The demonstration of the long-range behavior requires sampling the small-**q** regime on a non-uniform grid and separating the change of the exchange-correlation and self-energy contributions induced by phonon perturbations. In Fig. 4G and Fig. 4H, the agent successfully demonstrated the characteristic small-**q** divergence of the GW-level electron-phonon coupling and the corresponding divergent behavior from the many-body self-energy[39].

Finally, rigorous convergence is essential for reliable MBPT calculations. AI-Donkey therefore also encodes automated convergence workflows for GW and GW-BSE calculations. As a demonstration, we instruct the LLM agent to determine the computational parameters required to converge the quasiparticle band gap of monolayer $MoSe_2$. The LLM agent identified the coupled convergence of **k**-point sampling, number of empty bands, and dielectric cutoff, and iteratively updated these parameters until the target criterion is reached. Fig. 4I shows the band and dielectric-cutoff convergence at the converged **k**-point grid.

AI-Donkey represents workflows as modular combinations of skills and CLI, allowing LLM agents to execute complex, versatile, and customized workflows, such as evaluating self-energies at selected momenta and reusing intermediate quantities across related calculations. Furthermore, the modular design allows AI-Donkey to integrate advanced MBPT capabilities into LLM agent workflows. Features such as stochastic pseudobands[42] and patched sub-sampling of the Brillouin zone[43,44] are encoded natively in AI-Donkey as skills and CLIs, allowing them to be used by agents autonomously. For example, in Fig. 4, the agent applies stochastic pseudobands to speed up calculations of monolayer $MoSe_2$, graphene, and bulk diamond. A nonuniform sampling around the K and K' valleys are also used in monolayer $MoSe_2$ GW-BSE calculations to accelerate convergence. These features further reduce computational bottlenecks of MBPT calculations and are essential for making demanding GW and GW-BSE workflows affordable.

The capability of the underlying LLM remains important for long-horizon and non-standard workflow execution. AI-Donkey engineers large context windows to guide LLM generation, and the context size increases with the complexity of the workflow. The comparatively small context window size of the tested open-weight models, therefore, limits their performance on complex tasks. Additionally, invoking advanced features such as patched k-point sub-sampling deviates from the defined workflow protocols. In these non-standard workflows, AI-Donkey relies on the reasoning capabilities of the underlying LLM agent to assemble skill and CLI modules on-the-fly into dynamical workflows. Looking forward, AI-Donkey provides a general strategy for extending agentic automation beyond the specific GW, GW-BSE, and GWPT demonstrations presented in this work. The modular design of AI-Donkey will also allow recently developed advanced MBPT methods such as time-dependent adiabatic GW (TD-aGW)[45,46] and TD-aGW with exciton-phonon coupled dynamics[47], among others, to be integrated seamlessly.

## 3. Methods

### *3.1. AI-Donkey Implementation*

AI-Donkey interfaces with LLM agents through lightweight instruction files that define how the agent should interact with the repository. A `README.md` file provides environment setup instructions, while an `AGENTS.md` file routes the agent to predefined workflow protocols. After initialization, the LLM harness reads the appropriate workflow protocol, retrieves relevant skills and RAG documents, and invokes the CLI to generate inputs, link dependencies, validate consistency, and set up the calculation. The skill documents were curated by domain experts and organized by workflow module. Each skill is a markdown file containing the purpose of the module, required inputs and outputs, important physical parameters, common failure modes, and the corresponding CLI invocation. The RAG database was constructed from code documentation, workflow tutorials, and relevant literature, and was used to engineer the context of LLM agent workflows. The CLIs are implemented in Python and exposed as callable functions for individual workflow modules, and cover input generation, file linking, validation, and post-processing. AI-Donkey also includes HPC-specific skills and CLI utilities for job preparation and submission. In this work, AI-Donkey was run on Aurora at Argonne Leadership Computing Facility (ALCF) and Frontier at Oak Ridge Leadership Computing Facility (OLCF). The framework assumes that the required electronic-structure and MBPT codes are already compiled and available in the user environment, and that required external credentials, such as Materials Project API keys, are provided by the user. Additional HPC integrations can be added modularly without changing the core framework. This design allows AI-Donkey to be used by different agentic harnesses with minimal harness-specific adaptation. The AI-Donkey codebase was developed with the assistance of interactive coding agents (Claude Code).

### *3.2. Agentic Benchmarks*

We evaluated AI-Donkey using benchmark tasks designed to test whether LLM agents can construct complete MBPT workflows from high-level natural-language instructions. In each benchmark, the agent was run in a non-interactive session without human intervention. For Claude Code, we used the `claude -p` interface together with approval-bypass flags so that tool calls could be executed automatically during the benchmark. Codex and lightweight-harness runs were performed using equivalent non-interactive sessions with identical task prompts and knowledge-condition constraints. For each benchmark task, the same task prompt was used across all LLM agents and knowledge conditions, while system prompts were varied only to define the permitted knowledge access for each condition. A post-generation audit was carried out to ensure no unintentional access occurred during testing. Open-weight models benchmarked in this work are accessed through ALCF inference endpoints[48]. To evaluate open-weight models, we implemented a lightweight agentic harness consisting of a minimal agent loop and basic tool-calling templates, including shell execution, file reading, file writing, file searching, and directory listing. For the GWPT benchmark in Fig. 3, we used the same protocol with a shared task prompt across knowledge conditions and distinct system prompts to restrict access to the baseline, source-code, or full AI-Donkey setting. In GW, GW-BSE, and GWPT benchmarks in Fig. 2 and 3, one of the one-shot generations from each LLM agent and knowledge condition was submitted to HPC resources for runnability testing after generation. Additional details are provided in the supplementary material.

### *3.3. Scoring Rubrics*

Benchmark performance was evaluated using an expert-curated rubric defined separately for each task. Each rubric specifies the required workflow steps, key input flags, cross-stage dependencies, file-linking requirements, and format conversions needed for a valid workflow. An LLM judge assigned points by checking each rubric item iteratively and providing explicit justification with reference to the input files generated by the LLM agent. For Fig. 2 and 3, runnability and post-processing were evaluated separately as pass-or-fail based on successful program execution and generation of the required deliverables. To account for LLM stochasticity, static workflow generation was repeated five times for each benchmark condition and run-to-run sample standard deviation was computed from the resulting static workflow scores. Additional details are provided in the supplementary material.

### *3.4. Computational Workflows*

GW and GW-BSE workflows implemented in AI-Donkey use Quantum ESPRESSO[31] for DFT ground-state calculations and BerkeleyGW[10] for excited-state calculations. AI-Donkey also encodes a Wannier90[20] interface to obtain Wannier-interpolated quasiparticle band structure shown in Fig. 4A. The GWPT workflow was set up by AI-Donkey using ABINIT[15], BerkeleyGW, Wannier90, and EPW[21]. A symmetry unfolding interface, elph-interface, is also encoded in AI-Donkey to provide interfacing and post-processing capabilities[9]. For benchmarks in Fig. 2 convergence is not within the scope of the evaluation, and the LLM agents were given freedom to choose parameters. In Fig. 3, to keep the computational cost of the generated GWPT workflow reasonable, the main computational parameters such as **k**-point grid density and number of bands were specified in the task. In results presented in Fig. 4, we extract converged parameters from previous literature[40,41,49] and instruct LLM agents to use similar values in their calculations. The agent-generated workflows are submitted to HPC resources without manual modification of the physical parameters. HPC-related parameters and the submit script were sometimes adjusted as needed to address node failures, queue wait times, and calculation restarts. No scientific input files generated by the agent were manually edited for the results reported. However, we note that it is generally nontrivial to perform a proper Wannierization[50] and in this work, we focus on the MBPT execution and directly suggest effective parameters for Wannier90. This limitation can be mitigated in future work by interfacing AI-Donkey with automated Wannierization workflows[51]. Raw data used in post-processing and analysis in Fig. 4 are generated by non-interactive sessions. Wannier-interpolation in Fig. 4A was set up in an interactive Claude Code session using AI-Donkey. Real-space exciton wavefunction calculations are set up by non-interactive sessions using BerkeleyGW, in which the hole position is fixed, and the corresponding electron distribution is visualized. For monolayer $MoSe_2$, A- and B-series exciton character was determined by projecting band character using the eigenvectors of the BSE Hamiltonian. GWPT electron-phonon coupling quantities were extracted in an interactive Claude Code session using AI-Donkey. Additional details are provided in the supplementary material.

## 4. Conclusion

In this work, we developed AI-Donkey, an agentic framework that combines expert-curated knowledge with deterministic execution to enable high-fidelity end-to-end GW, GW-BSE, and GWPT calculations with LLM agents. We showed that current LLM agents, including frontier models, do not reliably construct MBPT workflows out of the box, averaging 54%, 36%, and 19% on GW, GW-BSE, and GWPT workflow benchmarks, with strong model dependence and large run-to-run variation. These failures largely originate from cross-stage inconsistencies, most notably in the handling of symmetry in k- and q-point grids, which propagate through the workflow, causing errors that are difficult to trace. This is particularly problematic in complex workflows where, providing access to the source code of the involved packages yields only limited improvement, highlighting the necessity of expert-curated knowledge for newly developed methods. With AI-Donkey, all tested LLM agents, from frontier proprietary models to open-weight models with a lightweight harness, construct correct, runnable workflows that reach full benchmark scores with no observed run-to-run variation. We further demonstrated calculations and post-processing of excited-state properties for 1D, 2D, and bulk materials, automated convergence studies, and the many-body enhancement of Fröhlich electron-phonon coupling in SiC. Owing to its modular design, AI-Donkey can be extended to additional codes and recently developed MBPT methods, broadening access to the predictive power of MBPT and providing a framework for agentic automation of complex, multi-code workflows.

## Supplementary material

The supplementary material is available and includes details on the implementation of AI-Donkey, benchmarks, and calculations performed in this work.

## Acknowledgments

This work is mainly supported by the United States (U.S.) Department of Energy (DOE), Office of Basic Energy Sciences, under the Contract No. DE-SC0026336. The work was partially supported by the U.S. National Science Foundation (NSF) under Grant No. OAC-2513830 in the development of interoperable software between the EPW and BerkeleyGW calculations, and by the Center for Computational Study of Excited-State Phenomena in Energy Materials (C2SEPEM) at Lawrence Berkeley National Laboratory (LBNL), which is funded by the U.S. DOE, Office of Science, Basic Energy Sciences, Materials Sciences and Engineering Division under Contract No. DE-AC02-05CH11231, as part of the Computational Materials Sciences Program, which provided advanced codes. B.L. and Y.W.C. acknowledge support from the National Research Foundation of Korea (NRF) under Grant No. RS-2024-00441954. C.E.H. and H.C.H. acknowledge support by National Science and Technology Council (NSTC) in Taiwan under Grants NSTC 114-2112-M-032-009-MY3 and 114-2811-M-032-012-MY2. An award of computer time was provided by the U.S. Department of Energy's (DOE) Innovative and Novel Computational Impact on Theory and Experiment (INCITE) Program. This research used supporting resources at the Argonne and the Oak Ridge Leadership Computing Facilities. The Argonne Leadership Computing Facility at Argonne National Laboratory is supported by the Office of Science of the U.S. DOE under Contract No. DE-AC02-06CH11357. The Oak Ridge Leadership Computing Facility at the Oak Ridge National Laboratory is supported by the Office of Science of the U.S. DOE under Contract No. DE-AC05-00OR22725.

**Author declarations**

The authors have no conflicts to disclose.

**Author Contributions**

Z.L. directed the research. B.Z., B.L., Z.Zhu., C.E.H., Z.Zheng., and M.D.B. developed the AI-Donkey framework and performed benchmarks. All authors discussed the results and contributed to the writing of the manuscript.

**Data availability**

Benchmark prompts, conversation logs, scoring rubrics, generated input files, and post-processing scripts are presented in the manuscript and supplementary material, and are available at https://github.com/Li-Group-USC/AI-Donkey-Paper-Artifacts-Release.git.

**Code availability**

The AI-Donkey repository, workflow protocols, skills, and CLI are open sourced at https://github.com/Li-Group-USC/AI-Donkey-Release.git. Standard releases of software codes used in this work are open source, with the latest versions available at https://www.abinit.org for ABINIT, http://www.wannier.org for Wannier90, https://berkeleygw.org for BerkeleyGW, https://www.quantum-espresso.org for Quantum ESPRESSO, and https://docs.epw-code.org for EPW.

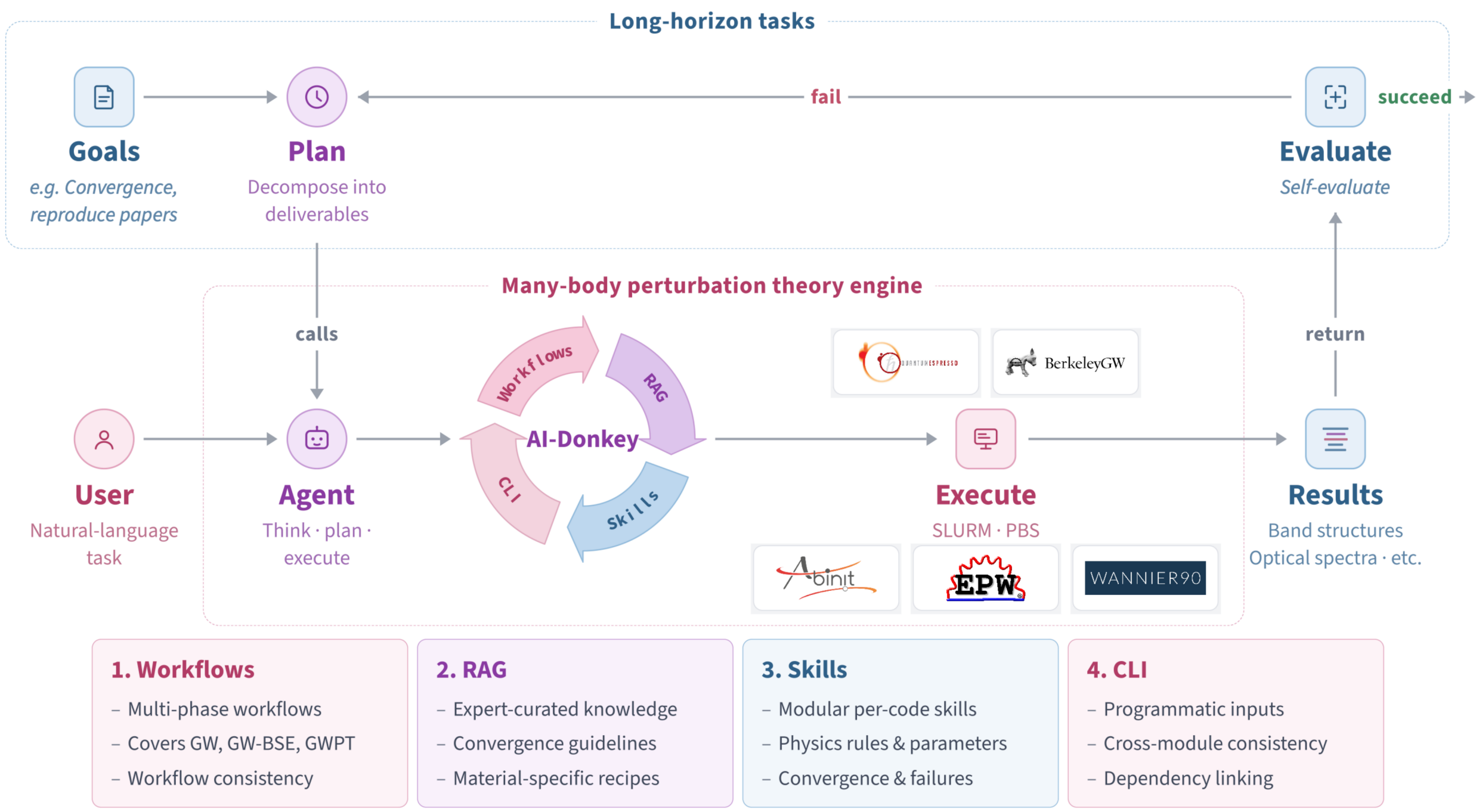

**Fig. 1. AI-Donkey architecture.** AI-Donkey provides a portable knowledge and execution layer for agentic MBPT calculations. Users use natural-language to define goals through an LLM harness, which invokes AI-Donkey to plan, execute, and verify GW, GW-BSE, and GWPT workflows. AI-Donkey consists of four components: workflow protocols that define ordered calculation steps; a RAG layer that retrieves expert-curated knowledge and guidance; modular skills that encode code-specific procedures; and a deterministic CLI layer for input generation, dependency linking, and execution. AI-Donkey is currently interfaced with Quantum ESPRESSO, ABINIT, BerkeleyGW, EPW, and Wannier90, and can be straightforwardly extended to other functionalities and packages.

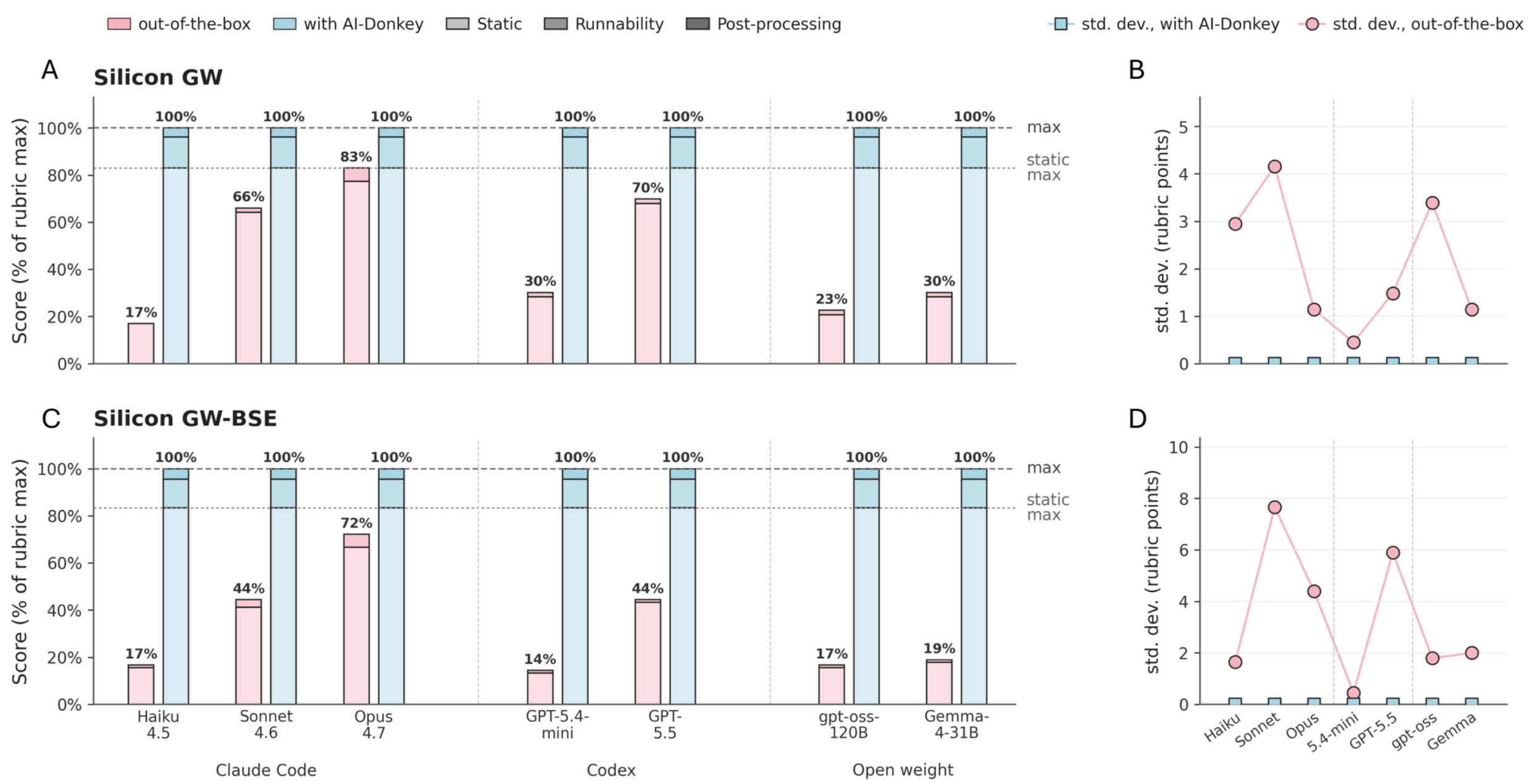

**Fig. 2. Benchmark on GW and GW-BSE tasks.** (A) Benchmarks of LLM agents on GW workflow for Si with and without AI-Donkey. Bars show percentage scores of a single one-shot run that is submitted to HPC resources for runnability testing, with a total score of 53 points. (B) Run-to-run sample standard deviation of the static workflow score across 5 repeated runs in rubric points. (C) Benchmarks of LLM agents on GW-BSE workflow for Si. Bars show percentage scores of a single one-shot run that is submitted to HPC resources for runnability testing, with a total score of 90 points. (D) Run-to-run sample standard deviation of the static workflow score across 5 repeated runs in rubric points.

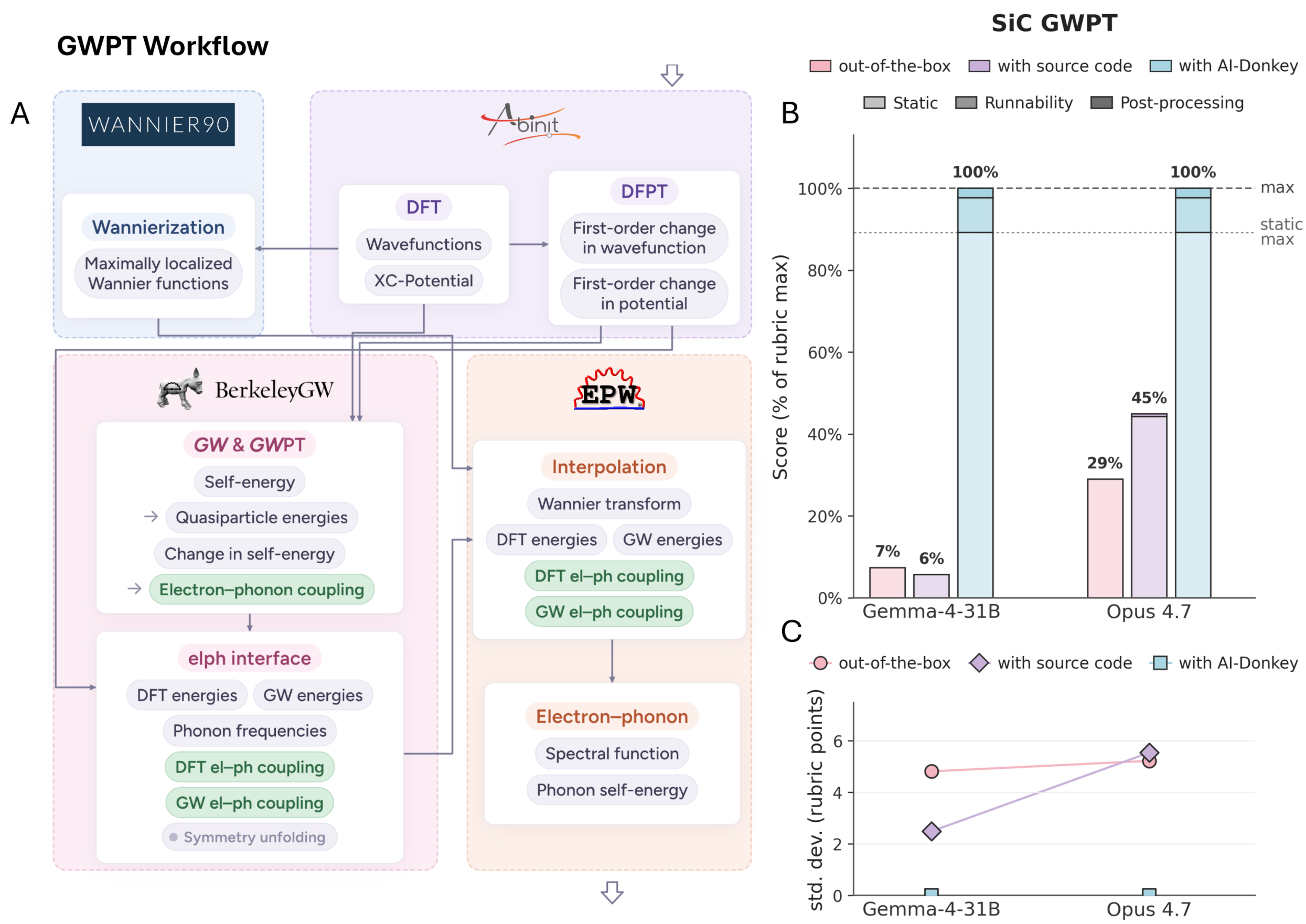

**Fig. 3. Benchmark on GWPT tasks.** (A) The GWPT workflow across ABINIT, Wannier90, BerkeleyGW, and EPW, which allows for electron-phonon coupling property calculations such as phonon-mediated superconductivity, mobility, electron spectral functions, and phonon linewidths, among others, at the full many-body level. The workflow is on electron linewidth of SiC, which consists of 20 individual computational steps and coupled physics parameters[9]. (B) Benchmark across 3 knowledge conditions using Opus 4.7 and Gemma-4-31B LLM agents, bars show the percentage scores for single one-shot generation that is submitted to HPC resources for runnability testing, with a total score of 176 points; and (C), run-to-run sample standard deviation of the static workflow score across 5 repeated runs in rubric points.

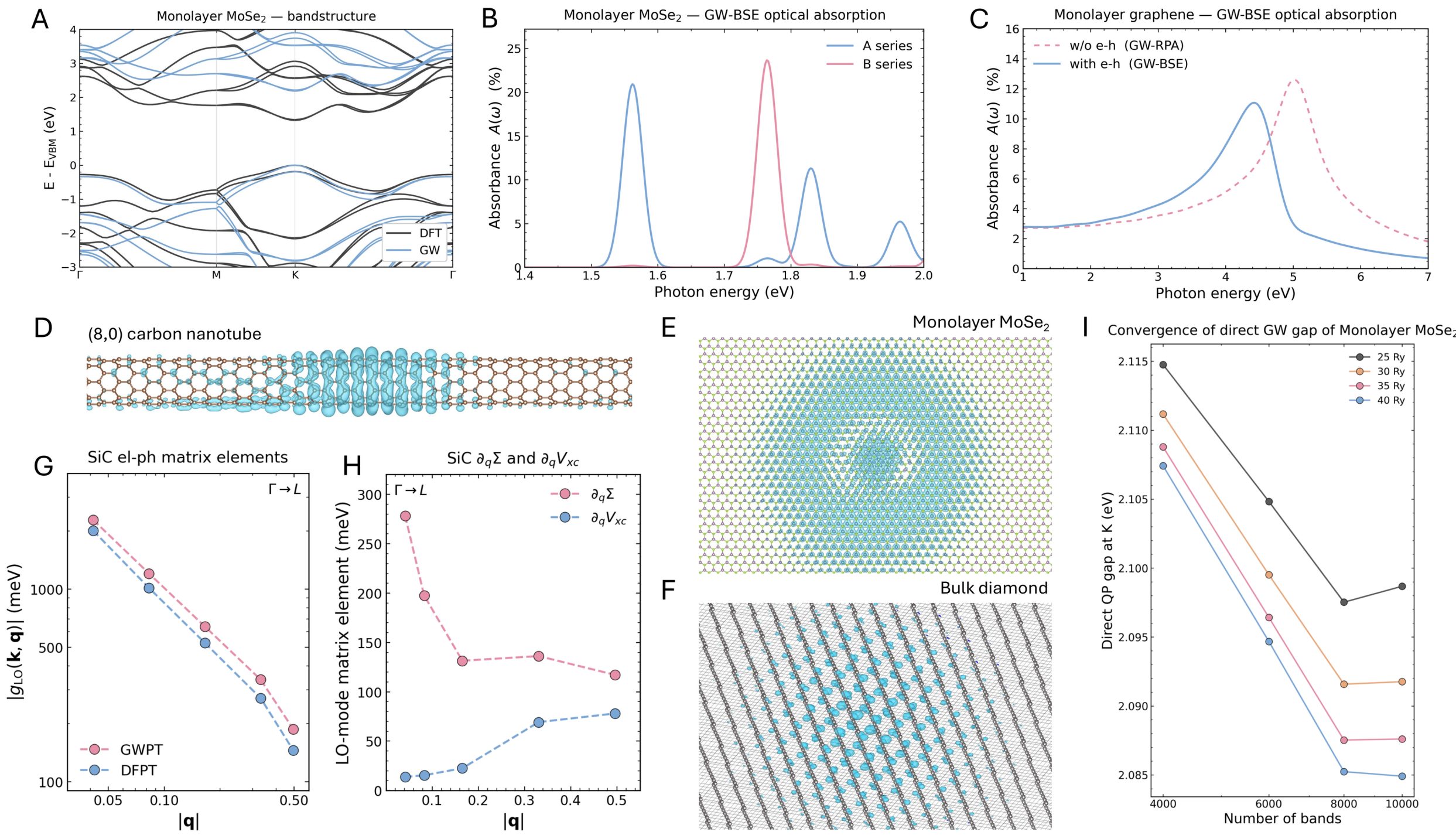

**Fig. 4. End-to-end MBPT calculations using AI-Donkey.** (A) Wannier-interpolated quasiparticle band structure of monolayer $MoSe_2$; (B) GW-BSE optical absorbance of monolayer $MoSe_2$ decomposed into spin-orbit-split A and B characters; (C) GW-BSE optical absorbance of monolayer graphene; (D-F) Real-space distribution of the electron part of the exciton wavefunction (with the hole position fixed) for (8,0) carbon nanotube, monolayer $MoSe_2$, and bulk diamond; (G) Magnitude of the GWPT-level and DFPT-level electron-phonon coupling matrix elements approaching the small-**q** limit, and (H) the divergent behavior of the phonon-induced changes in the exchange-correlation ($V_{XC}$) and self-energy ($\Sigma$) matrix elements; (I) Automated convergence test of the direct quasiparticle gap of monolayer $MoSe_2$.